\documentclass[10pt, prl,aps,twocolumn,preprintnumbers,superscriptaddress,amssymb,amsmath]{revtex4-1}
\usepackage{graphicx}
\usepackage[colorlinks=true, pdfstartview=FitV, linkcolor=red, citecolor=blue, urlcolor=blue]{hyperref}

\newcommand{\rmi}{\mathrm{i}}
\newcommand{\rme}{\mathrm{e}}
\newcommand{\rmd}{\mathrm{d}}
\newcommand{\kin}{\mathrm{kin}}
\newcommand{\EM}{\mathrm{EM}}
\newcommand{\tr}{\mathrm{tr}}

\newcommand{\bnabla}{\boldsymbol{\nabla}}
\newcommand{\bp}{{\boldsymbol{p}}}
\newcommand{\bq}{{\boldsymbol{q}}}
\newcommand{\bB}{\boldsymbol{B}}
\newcommand{\bE}{\boldsymbol{E}}
\newcommand{\bA}{\boldsymbol{A}}
\newcommand{\bL}{\boldsymbol{L}}
\newcommand{\bx}{\boldsymbol{x}}
\newcommand{\bv}{\boldsymbol{v}}
\newcommand{\br}{\boldsymbol{r}}
\newcommand{\bPi}{\boldsymbol{\Pi}}
\newcommand{\bOmega}{\boldsymbol{\Omega}}
\newcommand{\bj}{\boldsymbol{j}}

\newcommand{\calP}{\mathcal{P}}
\newcommand{\R}{\mathrm{R}}
\newcommand{\E}{\mathrm{E}}
\newcommand{\LLL}{\mathrm{LLL}}

\renewcommand{\c}{@{\hspace{0.5em}}c@{\hspace{0.5em}}}

\begin{document}

\title{Preponderant Orbital Polarization in Relativistic Magnetovortical Matter}

\author{Kenji Fukushima}
\affiliation{Department of Physics, The University of Tokyo, 7-3-1 Hongo, Bunkyo-ku, Tokyo, 113-0033, Japan}
\author{Koichi Hattori}
\affiliation{Zhejiang Institute of Modern Physics, Department of Physics, Zhejiang University, Hangzhou, Zhejiang 310027, China}
\affiliation{Research Center for Nuclear Physics, Osaka University, 10-1 Mihogaoka, Ibaraki, Osaka 567-0047, Japan}

\author{Kazuya Mameda}
\affiliation{Department of Physics, Tokyo University of Science, Tokyo 162-8601, Japan}
\affiliation{RIKEN iTHEMS, RIKEN, Wako 351-0198, Japan}

\preprint{RIKEN-iTHEMS-Report-24}

\begin{abstract}
We establish thermodynamic stability and gauge invariance in the magnetovortical matter of Dirac fermions under the coexistent rotation and strong magnetic field.
The corresponding partition function reveals that the orbital contribution to bulk thermodynamics preponderates over the conventional contribution from anomaly-related spin effects.
This orbital preponderance macroscopically manifests itself in the sign inversion of the induced charge and current in the magnetovortical matter, and can be tested experimentally as the flip of the angular momentum polarization of magnetovortical matter when the magnetic field strength is increased.
\end{abstract} 

\maketitle

{\it Introduction---}%
While the proper formulation of angular momentum was a cornerstone in quantum physics since the early twentieth century, its new facets continue to emerge in quantum many-body systems.
In quantum optics, the angular momentum of light is crucial for understanding optical vortices~\cite{PhysRevA.45.8185}.
In condensed matter physics, angular momentum underpins the quantum Hall effect~\cite{Prange1990}, topological insulators~\cite{Hasan:2010xy}, and the physics of Skyrmions~\cite{Nagaosa2013}.
Advancements in orbitronics have opened up novel streams for information processing in addition to spintronics~\cite{Manchon:2019}.
The recent high-energy physics is inseparable from angular momentum dynamics of quarks, gluons, and hadrons, which poses challenging problems investigated with relativistic colliders~\cite{Accardi:2012qut,*Leader:2013jra,*AbdulKhalek:2021gbh,STAR:2017ckg}.

A typical characteristic of angular momentum is its polarization under an external magnetic field or vorticity.
A lot of magnetic or rotational responses are empirically understood with the spin polarization.
Two of the prominent examples are the Einstein-de~Haas effect~\cite{einstein1915experimental} and the Barnett effect~\cite{Barnett1915}.
Also, the energy shift by the spin polarization implies that the chiral condensate is
stabilized in a magnetized system~\cite{Klevansky:1992qe}, and destabilized in a rotating system~\cite{Jiang:2016wvv}.
A similar argument can be applied to the Chandrasekhar-Clogston limit of a superconductor~\cite{chandrasekhar1962note,*Clogston:1962zz}.
Furthermore, the spin-polarization argument underlies the chirality-induced physics of relativistic fermions, such as the chiral magnetic and vortical effects~\cite{Kharzeev:2007jp,*Fukushima:2008xe,Son:2009tf,Landsteiner:2011cp}.

\begin{figure}
\centering
\includegraphics[width=1\columnwidth]{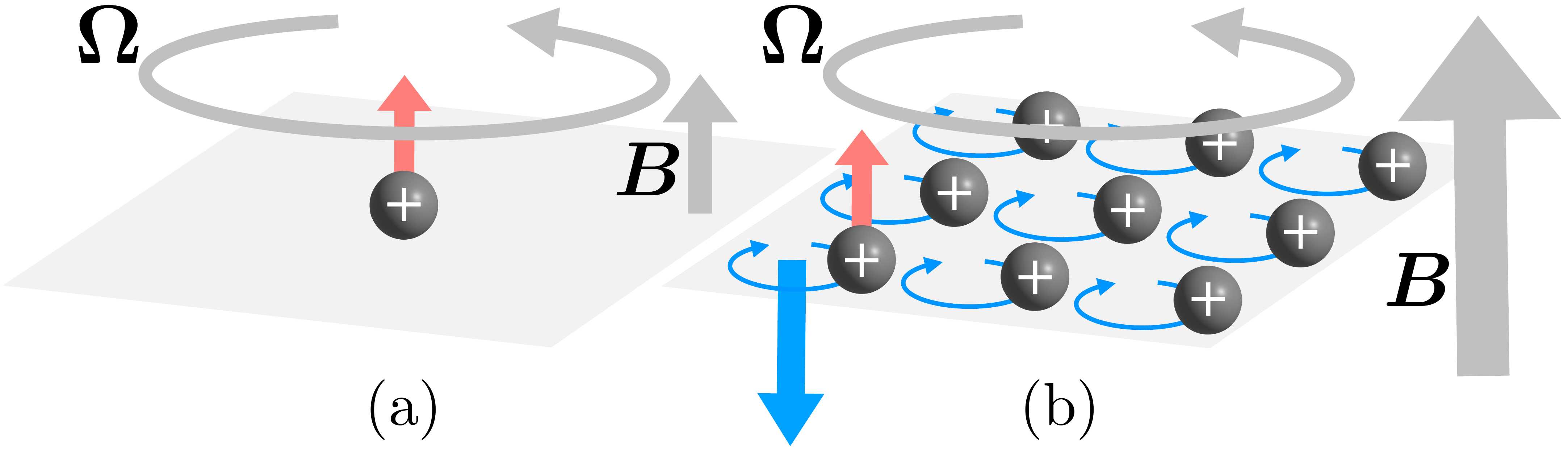}
    \caption{%
        Spin and orbital polarizations coupled with an external $\bOmega$ under (a) a weak $\bB$ and (b) a strong $\bB$.
        The orbital contribution in (a) is suppressed with widely spread cyclotron motion, while the orbital contribution surpasses the spin effects in (b).%
    }
    \label{fig:polarization}
\end{figure}

The following plausible argument is found for magnetovortical matter~\cite{Hattori:2016emy}.
While the Zeeman coupling distinguishes positive and negative charges, the spin-rotation coupling does not.
As a result, the interplay between a parallel magnetic field ($\bB=B\hat{z}$) and a vortical field ($\bOmega=\Omega\hat{z}$) provides an energy imbalance between positive and negative charges, and thermodynamically induces a nonvanishing net charge.
Such an intuitive picture has indeed been implemented with the linear response theory;
the charge density of chiral fermions under $\bB$ and $\bOmega$ is shown to be $\rho = eB\Omega/(4\pi^2)$~\cite{Hattori:2016njk}, which has also been reproduced with the chiral kinetic theory~\cite{Yang:2020mtz,*Mameda:2023ueq,*Yang:2024sfp,Lin:2021sjw}.
The effective-theory approach with an axial gauge field revealed the connection to a quantum anomaly and an associated induced current $\bj  =  e \bE \times\bOmega/(4\pi^2)$~\cite{Yamamoto:2021gts}. 
Owing to their nondissipative nature, the corresponding transport coefficients should be essential ingredients of the constitutive relations in (chiral) magnetohydrodynamics~\cite{Kharzeev:2011ds, *Hernandez:2017mch,*Hattori:2017usa,*Hongo:2020qpv,*Wang:2023imu}; see Ref.~\cite{Hattori:2022hyo} for a review.

This Letter aims to challenge the conventional spin-dominant argument for the polarization effects as sketched in Fig.~\ref{fig:polarization}~(a).
In magnetovortical matter, the crucial piece is the significant contribution from the orbital angular momentum  around the guiding center of each cyclotron orbit, as illustrated in Fig.~\ref{fig:polarization}~(b).
The polarization of the orbital angular momentum is opposite to that of the spin, as understood from Lenz's law.
In quantum theory, we show that this orbital polarization preponderates over the spin polarization in magnitude, and thus the direction of the total angular momentum is flipped.
Such a sign inversion definitely affects thermodynamic quantities;
for example, the charge density becomes 
\begin{equation}
\label{eq:charge-density}
 \rho = -\frac{eB\Omega}{4\pi^2},
\end{equation}
with an opposite sign to the aforementioned result. 

In this Letter, we formulate the thermodynamics of the magnetovortical matter, and derive the charge density [Eq.~\eqref{eq:charge-density}] and the associated spatial current with special attention to the overall sign.
We reveal that reaching the correct answer requires careful implementation of thermodynamic stability and gauge invariance, which were not encompassed in the previous thermodynamic function~\cite{Chen:2015hfc}.
These fundamental requirements turn out to be two sides of the same coin, and essential to resolving the unphysical divergence originating from large angular momenta~\cite{Ebihara:2016fwa,Fukushima:2020ncb}.
Besides, the thermodynamic function constructed in this way elucidates the relation among the family of the anomaly-related transport including Eq.~\eqref{eq:charge-density}.

{\it Angular momentum in the Landau-level basis---}%
In the quantum theory of charged particles, one of the fundamental concepts is the gauge invariance of kinematical variables.
For instance, the physical momentum of a charged particle is not $\bp=-\rmi\bnabla$ but $\bPi = \bp-e\bA$ with an external electromagnetic field $\bA$ and a charge $e$.
A similar distinction is found in angular momenta.
The gauge-invariant and thus physical quantity is not the \emph{canonical} angular momentum, $\bL = \bx\times\bp$, but the \emph{kinetic} angular momentum, $\bL_\kin = \bx\times\bPi$.
The difference between them may well be interpreted as the electromagnetic contribution. 
Indeed, in the symmetric gauge, the canonical angular momentum is a Noether conserved charge for the rotational symmetry and is related to the kinetic one as $\bL = \bL_\kin + \bL_\EM$ with $\bL_\EM = \bx\times(e\bA)$~\cite{PhysRevLett.113.240404}.
The gauge dependence in $\bL$ and $\bL_\EM$ cancels out.

Quantum mechanics under $\bB=B\hat{z}$ is conveniently described by the Landau-level basis, that is, $|n,m\rangle =  (a^\dag)^n (b^\dag)^m  |0,0\rangle/ \sqrt{n!m!}$, where $a = (\Pi_x + \rmi s\Pi_y)/\sqrt{2|eB|}$, $b = \sqrt{|eB|/2}(X - \rmi s Y)$, and $s=\mathrm{sgn}(eB)$.
Here, the coordinate variables, $X = x + \Pi_y/(eB)$ and $Y = y -\Pi_x/(eB)$, are the guiding center of the cyclotron motion, and they are constants of motion if further interactions are turned off. 
The longitudinal component of the kinetic angular momentum is represented as $L_\kin = x\Pi_y - y\Pi_x = \Lambda + \Delta$ with
\begin{equation}
\begin{split}
\label{eq:Lambda-Delta}
 &\Lambda = (x-X)\Pi_y - (y-Y)\Pi_x = - s(2a^\dag a +1) ,
\\
 &\Delta = X\Pi_y - Y\Pi_x = \rmi s(a^\dag b^\dag - ab).
\end{split}
\end{equation}
The derivation is reviewed in the Supplemental Material~\cite{supplemental}.
The diagonal operator $\Lambda$ quantifies the local orbital motion around the guiding center, while the purely off-diagonal $\Delta$ describes the global orbital motion around the origin of the coordinate system.
The form of $\Lambda$ can be understood as follows. 
For $eB>0$, the sign of the $\Lambda$ eigenvalue should be negative as the cyclotron motion is clockwise on the $xy$ plane.
The magnitude should be characterized by $a^\dag a$ 
so that higher Landau levels have larger $L_\kin$.
Also, $\Lambda$ as well as the energy levels are independent of $b$ and $b^\dag$, as a result of the translational symmetry that leads to the Landau degeneracy. 
Contrarily, $\Delta$ obviously breaks the translational symmetry, and thus does not play any role in characterizing conventional magnetized systems without rotation;
see also Ref.~\cite{Son:2024fgn}, in which $\Lambda$ enters the kinetic theory and the hydrodynamics describing the guiding center motion, while $\Delta$ does not.

{\it Gauge invariance in the partition function---}%
The traditional approach to installing the rotational effect in thermodynamics is based on the maximum entropy principle.
First, let us examine rigidly rotating matter (i.e., systems with rotational symmetry) in the absence of a gauge field.
Then, the angular velocity enters a statistical ensemble as the Lagrange multiplier for the angular momentum conservation, in parallel to those of inverse temperature and chemical potential for the conservation of energy and charge, respectively.
The partition function is thus given as~\cite{Landau:statistical1}
\begin{equation}
\label{eq:Z}
 Z = \tr\exp\Bigl[-\beta(\mathcal{H}-\Omega \mathcal{J})\Bigr]
 \quad
 \text{(no gauge field)}
\end{equation}
with the inverse temperature $\beta=1/T$ and the angular velocity $\boldsymbol{\Omega} = \Omega \hat{z}$.
Here, $\mathcal{H}$ and $\mathcal{J}$ are the field operators of a Hamiltonian and a conserved Noether charge, i.e., the $z$ component of the angular momentum in our setup, respectively.
Although the partition function [Eq.~\eqref{eq:Z}] was originally formulated in a nonrelativistic form, it yields the correct form of the chiral vortical effect~\cite{Vilenkin:1979ui,*Vilenkin:1980zv}, which was later confirmed with the linear response theory under a metric perturbation~\cite{Landsteiner:2011cp}. 

However, the partition function [Eq.~\eqref{eq:Z}] needs refinement for gauge theories.
The subtlety stems from the difference between the canonical and kinetic momenta elaborated above. 
Whereas the conserved $\mathcal{J}$ is built with the canonical angular momentum, $L = xp_y-yp_x$, it causes a problem that the partition function can become gauge dependent.
Instead, the correct partition function should read
\begin{equation}
\label{eq:Z_Jkin}
  Z = \tr\exp\Bigl[-\beta(\mathcal{H}-\Omega \mathcal{J}_\kin)\Bigr],
\end{equation}
where $\mathcal{J}_\kin $ is the field operator built with the kinetic angular momentum, $L_\kin = x\Pi_y - y\Pi_x$.
The partition function [Eq.~\eqref{eq:Z_Jkin}] is one particular form of the covariant density operator~\cite{Weldon:1982aq,*Becattini:2012tc}. 

Let us now derive the gauge-invariant partition function [Eq.~\eqref{eq:Z_Jkin}].
The derivation is based on the balance condition within electromagnetism~\cite{Landau:electrodynamics}.
Suppose that a particle with charge $e$ is rotating on the circle 
with a radius $r=(x^2+y^2)^{1/2}$ and an angular velocity $\boldsymbol{\Omega} = \Omega \hat{z}$. 
This particle has a velocity $\boldsymbol{v} = r\Omega \hat{\theta}$ along the azimuthal angle $\theta$ direction. 
If we apply a magnetic field $\bB = B \hat{z}$, the particle feels a radial drift force, $\boldsymbol{F}_\mathrm{drift} = e\bv\times\bB = eB\Omega \br$.
Then, the circular motion is no longer closed since the particle flows outwardly (inwardly) when $eB>0$ ($eB<0$).
The particle motion can reach a steady state only when an additional electric field offsets the radial drift force. 
Such an additional electric field is generated by an effective scalar potential $eA_0 = \Omega L_\EM$ involved in $\mathcal{J}_\kin$.
In the symmetric gauge $\bA=(A^1,A^2,A^3)=(-By/2,Bx/2,0)$, 
one finds that $e\bE = -e\bnabla A_0 = -e\bnabla(xA^2-yA^1) =-\boldsymbol{F}_\mathrm{drift}$.
In many-body physics, the stability of the circular motion has a close connection to equilibration of magnetovortical systems. 
That is, while a usual global equilibrium state allows for a global rotation and translation, the magnetovortical systems need to satisfy the additional balance condition along the radial direction (see Table~\ref{table:Hs}).
The covariant form of the local equilibrium condition in Ref.~\cite{Buzzegoli:2020ycf} also results in the present condition between $\bB$ and $\bE$ in its special case%
~\footnote{
The condition is written as $\partial_\mu \zeta + \beta^\lambda F_{\lambda\mu} = 0$, where $\zeta$ and $\beta^\mu$ are the chemical potential and fluid velocity (multiplied by the inverse temperature), respectively.
The condition $\bE = -B\Omega \br$ corresponds to $\zeta=0$, $\beta^\mu = \beta(1,\bOmega\times\bx)$, $F_{21}=-F_{12}=B$, and $F_{0i}=E^i$. %
}. 

\begin{table}
\caption{Difference of total Hamiltonians}
\begin{equation*}
{\renewcommand{\arraystretch}{1}
 \begin{array}{|\c||\c|\c|\c|} \hline
 \text{Total Hamiltonian} & \boldsymbol{F}_\mathrm{drift}\, & e\boldsymbol{E}  \, & \text{Electric forces}\\  \hline\hline
 \mathcal{H}-\Omega\mathcal{J} 
	& eB\Omega \br
 	& 0 
    & \text{Unbalanced}\\ \hline
\mathcal{H}-\Omega\mathcal{J}_\mathrm{kin}
	& eB\Omega \br
 	& -eB\Omega \br  
    & \text{Balanced}\\ \hline
 \end{array}
} 
\end{equation*}
\label{table:Hs}
\end{table}

Hereafter, we focus on relativistic magnetovortical matter, i.e., the rotating and magnetized relativistic medium 
composed of Dirac fermions.
As shown in the Supplemental Material~\cite{supplemental}, the partition function is given by Eq.~\eqref{eq:Z_Jkin} with
\begin{equation}
 \mathcal{H} - \Omega \mathcal{J}_\kin
 = \int\rmd^3 x\, \bar\psi\Bigl[-\rmi \gamma^i D_i + M - \Omega \gamma^0 (L_\kin + S)\Bigr]\psi,
  \label{eq:H_kin}
\end{equation}
where we defined $D_i = \partial_i +\rmi eA_i$, a mass $M$, and the spin operator $S = \rmi \gamma^1\gamma^2/2$~\footnote{%
    The same expression is obtained when the gauge field in the rotating coordinate $(x',y')$ is employed as $A_\mu' = (0,By'/2,-Bx'/2,0)$.
    This is different from Ref.~\cite{Chen:2015hfc}, which adopts $A_\mu' = (-B\Omega r'^2/2,By'/2,-Bx'/2,0)$.%
}.

{\it Orbital preponderance in the strong magnetic field limit---}%
A standard manner to evaluate Eq.~\eqref{eq:Z_Jkin} with the Hamiltonian [Eq.~\eqref{eq:H_kin}] is to diagonalize the Dirac operator.
However, the Landau-level basis is not the eigenstate of the Dirac operator due to $\Delta$.
In a technical perspective, the difficulty of the present eigenvalue problem is more obvious in the corresponding differential equation of motion in the symmetric gauge;  a potential term $\propto r^4 = (x^2+y^2)^2$ arises and the anharmonic oscillator problem must be solved.
Also the translational noninvariance due to $\Delta$ requires the careful treatment of the boundary effect in the cylindrical coordinate, and this inhomogeneity makes the analysis more complicated.

Nevertheless, in the lowest Landau level (LLL) approximation, we can obtain the analytic expression of Eq.~\eqref{eq:Z_Jkin} and illustrate the importance of the orbital motion.
Inserting the complete set of the Ritus basis for the spin-degenerate Landau levels $|n,m\rangle_\mathrm{R} = \calP_+|n,m\rangle  + \calP_-|n-1,m\rangle$ with $ \calP_\pm = \frac{1}{2}(1\pm 2 sS)$
(see, e.g., Refs.~\cite{Hattori:2020htm,Hattori:2023egw}), one can show that 
\begin{equation}
\begin{split}
\label{eq:lnZ_LLL}
 \ln Z = \tr \ln (\gamma^0\partial_\tau - \rmi\gamma^3\partial_z + M -\gamma^0 \nu),
\end{split}
\end{equation}
where we used Eq.~\eqref{eq:Lambda-Delta} and defined the rotation-induced effective chemical potential
\begin{equation}
\label{eq:nu}
\nu =  \Omega \langle L_\kin +S \rangle_\LLL = -\frac{s\Omega}{2}
\end{equation}
with $\langle O \rangle_\LLL = {}_\R\langle 0, m|O|0,m\rangle_\R$.

A crucial remark is that $\nu$ involves not only the spin-rotation coupling (spin polarization) but also the orbital-rotation coupling (orbital polarization).
This is why the conventional picture in Fig.~\ref{fig:polarization}~(a) is insufficient in the strong magnetic field case.
The negative sign in Eq.~\eqref{eq:nu} is due to the orbital contribution, $\nu_\mathrm{orb} = \Omega\langle L_\kin \rangle_\LLL = - s\Omega$ preponderating over the spin contribution, $\nu_\mathrm{spin} = \Omega\langle S \rangle_\LLL = s\Omega/2$. 
Such a sign inversion is a unique feature in the coupling to $\Omega$.
For instance, the LLL energy dispersion incorporates a similar combination, i.e., the magnetization coupling, but this vanishes for the LLL due to the $g$ factor:
$eB\langle L_\kin + 2S \rangle_\LLL = 0$.

The evaluation of Eq.~\eqref{eq:lnZ_LLL} is similar to that in the magnetized system at finite density.
After performing the Matsubara summation, the pressure is obtained as 
\begin{equation}
\begin{split}
\label{eq:P_LLL}
 P
 = \frac{|eB|}{2\pi}  \int\frac{\rmd p_z}{2\pi} 
 \left[ \epsilon
+ T\sum_{\eta = \pm} \ln \big(1+\rme^{-\beta(\epsilon-\eta\nu)}\big)
 \right]		
 , 		
\end{split}
\end{equation}
where $\epsilon = \sqrt{p_z^2+M^2}$.
The summation over $m$ in the Landau-level basis was taken as $\sum_m = |eB|R^2/2$ with the radial system size $R$, so that the usual Landau degeneracy is reproduced.
Equation~\eqref{eq:P_LLL} is derived without considering the boundary effects, which have to be taken into account in usual rotating systems because of the causality constraint $R \leq |\Omega|^{-1}$.
This is justified for bulk thermodynamics as long as we impose $\sqrt{|eB|}\gg R^{-1}$.
The physical meaning of the inequality is that the Landau wave function is localized too tightly to perceive the existence of the boundary~\cite{Ebihara:2016fwa}.

The pressure [Eq.~\eqref{eq:P_LLL}] implies a nontrivial property of the magnetovortical matter.  To see this,
let us specifically look into the zero temperature limit of Eq.~\eqref{eq:P_LLL}:
\begin{equation}
\label{eq:P_LLL_vac}
 P= \frac{|eB|\Omega^2}{16\pi^2}
 	\Biggl[
 		\sqrt{1-\alpha}-\alpha\ln \frac{1+\sqrt{1-\alpha}}{\sqrt{\alpha}}
 	\Biggr]\theta(1-\alpha),
\end{equation}
where $\alpha = (2M/\Omega)^2$ and $\theta(x)$ is the step function.
For $|\Omega|/2>M$, hence, the rotational effect is visible even at the vacuum.
This property emerges from the formation of the Landau levels that remain low-energy modes. 
Indeed, the rotational effect is unseen in a magnetic-free
vacuum, due to the infrared energy gap demanded by the causality constraint~\cite{Ebihara:2016fwa}.

{\it Charge density and angular momenta---}%
The negative sign in Eq.~\eqref{eq:nu} turns out to be crucial in charge-dependent thermodynamic quantities.
As a prominent example, let us compute the charge density in the massless limit $M\to 0$.
It is now convenient to add a real chemical potential $\mu$ through the shift $\Omega\mathcal{J}_\kin \to \Omega\mathcal{J}_\kin + \mu \int \rmd^3 x\, \bar{\psi}\gamma^0 \psi$ in Eq.~\eqref{eq:H_kin}, and accordingly $\nu \to \nu + \mu$ in Eq.~\eqref{eq:P_LLL}.
The pressure is then analytically evaluated as
\begin{equation}
\label{eq:P_mat}
 P = \frac{|eB|}{4\pi^2}
\left[  \biggl(\mu-\frac{s\Omega}{2}\biggr)^2
 + \frac{\pi^2T^2}{3} 
 \right].
\end{equation}
The charge density is calculated as $\partial P/\partial\mu = |eB|\mu/(2\pi^2)+ \rho$ with the magnetovortical contribution $\rho$ in Eq.~\eqref{eq:charge-density}.
The negative sign in $\rho$ originates in the same way as the effective chemical potential [Eq.~\eqref{eq:nu}].
Namely, the charge density $\rho$ is given as 
the sum of the spin and orbital contributions: 
\begin{equation}
 \rho_\mathrm{spin} = \frac{eB\Omega}{4\pi^2},
 \qquad
 \rho_\mathrm{orb} = -\frac{eB\Omega}{2\pi^2}.
\end{equation}

We make comparisons with related studies.
First, in Refs.~\cite{Ebihara:2016fwa,Fukushima:2020ncb}, we perform a similar analysis with the partition function [Eq.~\eqref{eq:Z}], instead of Eq.~\eqref{eq:Z_Jkin}.
Then, $\rho_\mathrm{spin}$ was correctly found, but the orbital contribution corresponding to $\rho_\mathrm{orb}$ was divergent in the large $R$ limit, as $\simeq eB\Omega N/(4\pi^2)$ with the magnetic flux $N=eBR^2/2 \gg 1$~\footnote{%
    The derivation is done almost parallelly to the present case, by replacing $\langle L_\kin\rangle_\LLL=-s$ with $\langle L\rangle_\LLL = sm$, which follows from the ladder-operator form in the symmetric gauge, $L=xp_y-yp_x = b^\dag b-a^\dag a$. 
    This leads to the thermal average with $\sum_{m=0}^N \langle L\rangle_\LLL \simeq sN^2/2$, instead of $\sum_{m=0}^N \langle L_\kin\rangle_\LLL = -sN$.
    Such an extra factor of $N$ results in the divergent orbital part of the charge density.%
}.
This unphysical divergence is identified as the thermodynamic instability because of the nonvanishing drift force in the partition function [Eq.~\eqref{eq:Z}].
Second, the Kubo formula based on the diagrammatic approach led to the answer that looked like only $\rho_\mathrm{spin}$~\cite{Hattori:2016njk}. 
The sign of $\rho$ was skewed due to the lack of clear recognition of the orbital-rotation coupling.
As we discuss in the Supplemental Material~\cite{supplemental}, the careful analysis of the linear response theory agrees with Eq.~\eqref{eq:charge-density}~\footnote{%
The linear response theory under a slow rotation should be equivalent to the analysis from the partition function. On the other hand, this is not the case under a perturbation not realizing an equilibrium state, as is found in the chiral magnetic conductivity under an ac magnetic field~\cite{Kharzeev:2009pj}.%
}.

Besides, the chiral kinetic theory (CKT) up to $O(\hbar^2)$ also reproduces only the spin part $\rho_\mathrm{spin}$~\cite{Yang:2020mtz,*Mameda:2023ueq,*Yang:2024sfp}.
This, however, does not contradict our analysis in the LLL approximation.
In the usual CKT, background electromagnetic fields are assumed to be small compared with other scales.
Since a weak magnetic field cannot yield the Landau quantization~\cite{Chen:2017xrj}, only the spin-rotation coupling contributes to the induced charge density [see Fig.~\ref{fig:polarization}~(a)].
On the other hand, the CKT constructed with the Landau-level basis~\cite{Lin:2019fqo} should pick up the orbital contribution $\rho_\mathrm{orb}$.
Our result appears to be consistent with the ``total'' charge density obtained from the ``vortical solution'' in Ref.~\cite{Lin:2021sjw}.

The above observations elucidate the distinct origins of $\rho_\mathrm{spin}$ and $\rho_\mathrm{orb}$. 
The former origin persists irrespective of the magnitude of the magnetic field. 
The latter emerges only under a strong magnetic field. 
Hence, it is deduced that only $\rho_\mathrm{spin}$ is related to a quantum anomaly.
We confirm this identification with the thermodynamic potential.
From the pressure [Eq.~\eqref{eq:P_mat}], the total angular momentum is derived as $\partial P/\partial\Omega = J_\mathrm{spin} + J_\mathrm{orb} + |eB|\Omega/(8\pi^2)$, where the last term is the contribution from the effective chemical potential $\nu=-s\Omega/2$ and the others are from the real one $\mu$: 
\begin{equation}
\begin{split}
\label{eq:J}
 J_\mathrm{spin} = \frac{eB\mu}{4\pi^2},
 \quad
 J_\mathrm{orb} = -\frac{eB\mu}{2\pi^2}.
\end{split}
\end{equation}
The Maxwell relation for $\partial^2 P/\partial\mu\partial\Omega$ shows that $\rho_\mathrm{spin}$ and $J_\mathrm{spin}$ share the same origin of the coefficient $eB/4\pi^2$; see also a similar correspondence in Ref.~\cite{Huang:2017pqe}.
We recall that the fermionic spin is also connected with the axial current as $J_\mathrm{spin} = j^z_5/2$.
Therefore, both $\rho_\mathrm{spin}$ and $J_\mathrm{spin}$ are anomaly related, and so is the chiral separation current $j^z_5 = eB\mu/(2\pi^2)$.
We note that they receive the mass correction, which is again the same as the chiral separation effect.

Finally, we discuss an associated transport phenomenon in magnetovortical matter.
When the induced charge density $ \rho$ is 
rotating with the velocity $ v_\theta = r \Omega$, the azimuthal current density is given as $j_\theta = \rho  v_\theta  =  - eB\Omega^2 r/(4\pi^2)$.
The stability condition for electromagnetic fields, $\bE=-B\Omega\br$, leads to the vector form $\bj = -e \bE \times\bOmega/(4\pi^2)$, which agrees with the spatial component of the Chern-Simons current~\cite{Yamamoto:2021gts}, but with the opposite sign for the same reason as $\rho$. 
This explains the physical picture behind the Chern-Simons current derived from the discussion about quantum anomaly. 

{\it Outlook---}%
Our work sheds light on the orbital angular momentum carried by magnetized Dirac fermions. 
Lattice QCD simulation is a first-principles method to extend our analysis to the strongly coupled magnetovortical matter with dynamical gluons. 
A benchmark quantity is magnetization, which has been simulated without rotation~\cite{Bali:2012jv,Bali:2013owa,*Bonati:2013lca,*Levkova:2013qda,*Bali:2020bcn}. 
In the magnetovortical matter, magnetization is obtained from our pressure [Eq.~\eqref{eq:P_mat}] as 
\begin{equation}
\begin{split}
\label{eq:M}
 \frac{\partial P}{\partial(eB)} 
 = \frac{s}{4\pi^2} \biggl(\mu^2 - s\mu\Omega + \frac{\Omega^2}{4}
 +\frac{\pi^2T^2}{3} \biggr).
\end{split}
\end{equation}
The sign of the term $\propto \mu\Omega$ provides a crucial test.
Our analysis is also inspiring enough to revisit the phase structures of magnetovortical matter~\cite{Chen:2015hfc,Liu:2017spl,*Cao:2019ctl,*Chen:2019tcp} with the correct thermodynamic formulation in lattice QCD\@. 
In general, numerical simulations for rotating systems depend on details of boundary conditions implemented for causality~\cite{Yamamoto:2013zwa,*Braguta:2021jgn,*Yang:2023vsw}. 
Investigating the magnetovortical matter is technically advantageous in this respect because the wave functions in strong magnetic fields are tightly localized within a cyclotron radius. 

Our finding in Eq.~\eqref{eq:J} at $\Omega = 0$ implies that the strongly magnetized matter plays a role of the storage of the negative angular momentum $J = J_\mathrm{spin} + J_\mathrm{orb} <0$ for $eB\mu>0$.
If we adiabatically turn off $B$, a stored $J$ is distributed in a rigid body so that the total angular momentum is conserved.
This evokes the Einstein-de Haas (EdH) effect, but the rotation occurs in an opposite direction to the conventional one, which we call the anti-EdH effect.
This novel phenomenon can be tested in parallel to the conventional one by using Weyl or Dirac semimetals, which are good environments for the direct applicability of the present relativistic analysis, the avoidability of the mass correction, and the controllability of the chemical potential~\cite{PhysRevB.108.115142}.
The maximum induced angular momentum is evaluated from $\Omega_\mathrm{ind} = |J|/I$ with $I$ being the density of moment of inertia for a cylindrical material. 
The estimated magnitude is as large as
\begin{equation}
 \begin{split}
 \Omega_\mathrm{ind} 
  = \frac{\displaystyle \biggl(\frac{\mu}{10^{-1}\,\text{eV}}\biggr)\biggl( \frac{B}{10^2\,\text{T}}\biggr)}{\displaystyle \biggl(\frac{\rho_\mathrm{mass}}{1\,\text{g}/\text{cm}^3}\biggr) \biggl(\frac{R}{10^{-2}\,\text{cm}}\biggr)^2} 
  \times 7.76\times 10^{-5}\, \text{rad/s},
 \end{split}
\end{equation}
where $\rho_\mathrm{mass}$ is the mass density and the Fermi velocity is introduced as $v_\mathrm{F} = 10^6\,\text{m}/\text{s}$.
This $\Omega_\mathrm{ind}$ is smaller by only 1 order of magnitude than the conventional EdH, and sufficiently testable;
for example, a realistic set of parameters $\mu=10^{-1}\,\text{eV}$, $B = 10^2\,\text{T}$, and $R=10^{-2}\,\mathrm{cm}$ leads to $\mu/\sqrt{2\hbar v_\mathrm{F}^2 eB}=0.39$ and $eBR^2/2\hbar =1.52\times 10^{9}$, which justify the LLL approximation with the large degeneracy, and thus the estimate based on Eq.~\eqref{eq:J}.
Further studies open up a new avenue to the orbitronics and related phenomena of Dirac electrons. 

\begin{acknowledgments}
{\it Acknowledgments---}%
The authors thank Takuya~Shimazaki for his contribution to the early stage of this work, and Matteo~Buzzegoli, Hao-Lei~Chen, Maxim~Chernodub, Masaru~Hongo, Xu-Guang~Huang, Shu~Lin, Xin-Li~Sheng, Igor~Shovkovy, Dam~Thanh~Son, Qun~Wang, Naoki~Yamamoto, Di-Lun~Yang, and Yi~Yin for valuable comments and discussions.
The authors also appreciate the hospitality of ECT* during the workshop ``Spin and Quantum Features of QCD Plasma,'' where this work was discussed and finalized.
This work is supported by the Japan Society for the Promotion of Science KAKENHI Grant No.~20K03948, No.~22H01216, No.~22H05118, and No.~24K17052.
\end{acknowledgments}

\bibliographystyle{apsrev4-2}
\bibliography{rotation}


\clearpage
\onecolumngrid

\begin{center}
        \textbf{\large --- Supplemental Material ---\\ $~$ \\
        Preponderant Orbital Polarization in Relativistic Magnetovortical Matter}\\
        \medskip
        \text{Kenji~Fukushima, Koichi~Hattori, and Kazuya~Mameda}
\end{center}
\setcounter{equation}{0}
\setcounter{figure}{0}
\setcounter{table}{0}
\setcounter{page}{1}
\makeatletter
\renewcommand{\thesection}{S\arabic{section}}
\renewcommand{\theequation}{S\arabic{equation}}
\renewcommand{\thefigure}{S\arabic{figure}}
\renewcommand{\bibnumfmt}[1]{[S#1]}

\section{Landau quantization}
\label{app:landau}
This section is devoted to a review of the nonrelativistic Landau problem (see also Ref.~\cite{Hattori:2023egw});
the following argument is also applied to the energy dispersion of Dirac fermions.
The one-particle Hamiltonian under a magnetic field $\bB=B\hat{z}$ is given by
\begin{equation}
\label{eq:Hqm}
 H = \frac{1}{2M}(\Pi_x^2 + \Pi_y^2),
 \qquad
 \Pi_i = p_i-eA_i (x,y) ,
\end{equation}
where $A_i$ is the gauge field such that $\partial_x A_y-\partial_y A_x = B$.
Here $\Pi_i$ is called the kinetic (linear) momentum;
`kinetic' means this is the gauge-invariant and thus physical momentum.
Another fundamental quantity to characterize the cyclotron motion is $(X,Y)$ defined as
\begin{equation}
\label{eq:XY}
 X = x + \frac{\Pi_y}{eB}  , \qquad
 Y = y - \frac{\Pi_x}{eB}  .
\end{equation}
From a solution of the classical equation of motion, $\dot{p}_i=-\partial H/\partial x_i$ and $\dot{x}_i=\partial H/\partial p_i$, one finds that the set of $(X,Y)$ is the center coordinate of the cyclotron motion, which is referred to as the guiding center coordinate.

An important advantage of introducing the guiding center $(X,Y)$ is that with the help of the canonical commutation relation $[x_i,p_j] = \rmi \delta_{ij}$, one can construct the following algebra: 
\begin{equation}
  [\Pi_x,\Pi_y] = \rmi eB  ,
  \qquad
  [X,Y] = -\frac{\rmi}{eB}  , 
  \qquad
  [X,\Pi_i] = [Y,\Pi_i] = 0.
\end{equation}
This enables us to introduce the following two types of the ladder operators:
\begin{equation}
\label{eq:ab}
 a := \frac{1}{\sqrt{2|eB|}} (\Pi_x + \rmi s\Pi_y)  , \qquad
 b := \sqrt{\frac{|eB|}{2}} (X - \rmi s Y)  ,
 \qquad
 s:=\mathrm{sgn}(eB),
\end{equation}
which satisfies $[a,a^\dag] = [b,b^\dag]=1$ and $[a,b] = 0$.
We inversely solve Eq.~\eqref{eq:ab} in terms of $\Pi_i$, $X$ and $Y$ as
\begin{equation}
 \Pi_x = \sqrt{\frac{|eB|}{2}}(a+a^\dag)  , \qquad
 \Pi_y = -\rmi s\sqrt{\frac{|eB|}{2}}(a-a^\dag)  ,\qquad
 X = \frac{1}{\sqrt{2|eB|}}(b+b^\dag)  , \qquad
 Y = \frac{\rmi s}{\sqrt{2|eB|}}(b-b^\dag)  .
\end{equation}
The Landau-level basis is defined as the eigenstate for $a^\dag a$ and $b^\dag b$:
\begin{equation}
 |n,m\rangle :=  \frac{(a^\dag)^n (b^\dag)^m }{\sqrt{n!m!}} |0,0\rangle ,
\end{equation}
where the ground state is normalized as $\langle 0,0 | 0,0 \rangle = 1$.

Let us explain the physical meaning of the quantum numbers $n$ and $m$.
It is straightforward to show
\begin{equation}
 \Pi_x^2 + \Pi_y^2 = |eB|(2a^\dag a+1).
\end{equation}
This representation implies that $n$ corresponds to the transverse momentum on the $xy$-plane, or equivalently, to the quantum number to specify the Landau level, as is clear from
\begin{equation}
 H|n,m\rangle = \frac{|eB|}{M} \biggl(n + \frac{1}{2} \biggr) |n,m\rangle.
\end{equation}
An important observation here is that the Landau level is degenerate in terms of $m$.
The reason is understandable if we notice the following relation:
\begin{equation}
 X^2 + Y^2 = \frac{1}{|eB|} (2b^\dag b+1).
\end{equation}
This indicates that $m$ characterizes the guiding center position, which is obviously not invariant under the translation on the $xy$-plane.
The degeneracy of $m$ hence reflects the translational invariance of the Landau problem.
More precisely, each state $|n,m\rangle$ corresponds to the cyclotron motion where the magnitude of the momentum is $\sqrt{|eB|(2n+1)}$ and the center position is located at the distance $\sqrt{(2m+1)/|eB|}$ from the origin.

Using the above ladder operators, we represent the kinetic angular momentum as
\begin{equation}
\begin{split}
 L_z^\kin 
 &:= x \Pi_y - y \Pi_x = \Lambda + \Delta,
\end{split}
\end{equation}
where $\Lambda$ and $\Delta$ are the diagonal and off-diagonal operator defined as
\begin{eqnarray}
 &&\Lambda:=(x-X)\Pi_y - (y-Y)\Pi_x 
 = - s(2a^\dag a +1) ,\\
 &&\Delta := X\Pi_y - Y\Pi_x 
 = -\rmi s(ab-a^\dag b^\dag) .
\end{eqnarray}
The operator $\Lambda$ corresponds to the usual cyclotron motion, i.e., the local circular motion around the guiding center.
The overall sign of $\Lambda$ is met to the classical Lenz's law, and it is plausible that the magnitude is proportional to the transverse momentum.
Another piece $\Delta$ is regarded as the angular momentum due to the global circular motion around the origin.
For this reason, $\Delta$ is nothing to do with the usual Landau problem respecting the translational symmetry, while it generally contributes under rotation, which washes out the translation invariance.
We emphasize that in the lowest Landau level approximation, the off-diagonal operator $\Delta$ does not enter the thermodynamics of magnetovortical matter.
This is physically because the strong magnetic field limit effectively recovers the translation invariance even under rotation.

\section{Field operator of angular momentum}
\label{app:angular}

We present the brief derivation of the gauge-invariant angular momentum operator composed of Dirac fields (see also Ref.~\cite{Fukushima:2020qta} for more detailed discussion), providing Eq.~(5) in the main text.
We start from the following Lagrangian:
\begin{equation}
  \mathcal{L} = \bar{\psi} ( \rmi \gamma^\mu D_\mu-m )\psi - \frac{1}{4} F_{\mu\nu} F^{\mu\nu} .
\end{equation}
with $D_\mu \psi = (\partial_\mu +\rmi eA_\mu)\psi$ and $F_{\mu\nu}=\partial_\mu A_\nu-\partial_\nu A_\mu$.
Here and hereafter the raising and lowering of indices are performed using the Minkowski metric $\eta_{\mu\nu} =\mathrm{diag}(1,-1,-1,-1)$.
The canonical energy-momentum tensor is computed as
\begin{equation}
\begin{split}
 \Theta^{\mu\nu} 
 &= \frac{\partial\mathcal{L}}{\partial(\partial_\mu \psi)}\partial^\nu\psi
    + \partial^\nu\bar\psi \frac{\partial\mathcal{L}}{\partial(\partial_\mu \bar\psi)}
    + \frac{\partial \mathcal{L}}{\partial(\partial_\mu A_\rho)}\partial^\nu A_\rho 
    - \eta^{\mu\nu}\mathcal{L} \\
 &= \bar\psi \rmi\gamma^\mu \partial^\nu \psi
    - F^{\mu\rho}\partial^\nu A_\rho
    - \eta^{\mu\nu}\mathcal{L}.
\end{split}
\end{equation}

The above energy-momentum tensor is not gauge-invariant obviously, but the situation can be cured through the Belinfante prescription;
the energy momentum tensor can be shifted by $\partial_\rho B^{\rho\mu\nu}$ such that $B^{\mu\rho\nu}=-B^{\rho\mu\nu}$ because the conservation law still holds as $\partial_\mu (\Theta^{\mu\nu}+\partial_\rho B^{\rho\mu\nu}) = 0$.
One of such Belinfante tensors above is
\begin{equation}
 B^{\rho\mu\nu} 
 = \frac{1}{2} (M_\mathrm{s}^{\rho\mu\nu}-M_\mathrm{s}^{\mu\rho\nu}+M_\mathrm{s}^{\nu\mu\rho})
 \qquad
 M_\mathrm{s}^{\mu\rho\sigma} 
 := \bar{\psi}\gamma^\mu\Sigma^{\rho\sigma}\psi
  + F^{\mu\rho}A^\sigma-F^{\mu\sigma}A^\rho,
\end{equation}
where $\Sigma^{\mu\nu}=(\rmi/4)[\gamma^\mu,\gamma^\nu]$.
This $M_\mathrm{s}^{\mu\rho\sigma}$ is a spin-like part of the total angular momentum tensor, as we will show below.
Thanks to the identity $\gamma^\mu \gamma^\nu \gamma^\rho=\eta^{\mu \nu} \gamma^\rho+\eta^{\nu \rho} \gamma^\mu-\eta^{\rho \mu} \gamma^\nu- \rmi \varepsilon^{\mu \nu \rho \sigma} \gamma^5 \gamma_\sigma$ with $\gamma_5 = \rmi\gamma^0\gamma^1\gamma^2\gamma^3$ and $\varepsilon^{0123}=1$, the Belinfante term is written as
\begin{equation}
 B^{\rho\mu\nu}
 = \frac{\rmi}{2} \bar\psi \biggl(\eta^{\nu\rho}\gamma^\mu-\eta^{\nu\mu}\gamma^\rho-\frac{\rmi}{2}\varepsilon^{\rho\mu\nu\sigma}\gamma_5\gamma_\sigma\biggr)\psi
 - F^{\rho\mu}A^\nu.
\end{equation}
Then, the improved energy-momentum tensor is 
\begin{equation}
\begin{split}
 T^{\mu\nu}
 &:=\Theta^{\mu\nu} + \partial_\rho B^{\rho\mu\nu}\\
 &=\bar\psi \rmi\gamma^\mu \overleftrightarrow{D^\nu} \psi 
    + \frac{1}{4}\varepsilon^{\mu\nu\rho\sigma}\partial_\rho(\bar\psi\gamma_5\gamma_\sigma\psi)
    - F^{\mu\rho} {F^\nu}_\rho 
    + \frac{1}{4} \eta^{\mu\nu}F_{\rho\sigma}F^{\rho\sigma} ,
\end{split}
\end{equation}
where we define $\overleftrightarrow{D_\mu} := (D_\mu-\overleftarrow{D_\mu})/2$ with $\bar\psi\overleftarrow{D_\mu} := \bar\psi(\overleftarrow{\partial_\mu}-\rmi eA_\mu)$.
In deriving the last line, we utilize the U(1) current conservation $\partial_\mu (\bar\psi\gamma^\mu\psi)=0$ and the equations of motions $(\rmi\gamma^\mu D_\mu-m) \psi =0$ and $\partial_\mu F^{\mu\nu}=e\bar\psi\gamma^\nu\psi$.

Similarly, the canonical angular momentum tensor is derived as
\begin{equation}
\begin{split}
 M^{\lambda\mu\nu} 
 &= x^\mu \Theta^{\lambda\nu} - x^\nu \Theta^{\lambda\mu} 
 - \rmi
 \biggl[
    \frac{\partial\mathcal{L}}{\partial (\partial_\lambda \psi)} \Sigma^{\mu\nu}\psi
    + \bar\psi \Sigma^{\mu\nu} \frac{\partial\mathcal{L}}{\partial (\partial_\lambda \bar\psi)}
    + \frac{\partial\mathcal{L}}{\partial (\partial_\lambda A_\rho)} \Xi^{\mu\nu\rho\sigma} A_\sigma
 \biggr]\\
 &= x^\mu \Theta^{\lambda\nu} - x^\nu \Theta^{\lambda\mu} + M_\mathrm{s}^{\lambda\mu\nu}.
\end{split}
\end{equation}
with $\Xi^{\mu\nu\alpha\beta} = \eta^{\mu\alpha}\eta^{\nu\beta}-\eta^{\mu\beta}\eta^{\nu\alpha}$.
Again the above form is not gauge-invariant, but can be improved by the Belinfante tensor;
the angular momentum can be shifted by $\partial_\rho(x^\mu B^{\rho\lambda\nu}-x^\nu B^{\rho\lambda\mu})$ because $B^{\lambda\rho\nu}=-B^{\rho\lambda\mu}$ ensures the conservation law $\partial_\lambda [ M^{\lambda\mu\nu} + \partial_\rho(x^\mu B^{\rho\lambda\nu}-x^\nu B^{\rho\lambda\mu})] = 0$.
From $B^{\mu\lambda\nu}-B^{\nu\lambda\mu}=-M_\mathrm{s}^{\lambda\mu\nu}$, we find
\begin{equation}
\begin{split}
 J^{\lambda\mu\nu} 
 &:= M^{\lambda\mu\nu} + \partial_\rho(x^\mu B^{\rho\lambda\nu}-x^\nu B^{\rho\lambda\mu})\\
 &= x^\mu T^{\lambda\nu} - x^\nu T^{\lambda\mu}.
\end{split}
\end{equation}

The conserved energy density and angular momentum density along the $z$-axis are ${T^0}_0$ and  $J^{0xy}$, respectively.
Keeping only the fermionic contributions, we arrive at the following expressions of the Hamiltonian and the angular momentum:
\begin{equation}
\begin{split}
 \mathcal{H}
 &:= \int\rmd^3 x \, {T^0}_0
 = \int\rmd^3 x \,\bar\psi (-\rmi\gamma^i D_i - m)\psi
 ,\\
 \mathcal{J}_\mathrm{kin} 
 &:= \int\rmd^3 x \, J^{0xy}
 = \int\rmd^3 x\, \bar\psi \gamma^0 \biggl[-\rmi(xD_y-yD_x) + \frac{\rmi}{2}\gamma^1\gamma^2\biggr]\psi ,
\end{split}
\end{equation}
where we utilize the Dirac equation and perform the integration by parts.
These gauge-invariant operators correspond to those in Eq.~(5) in the main text.

\section{Linear response theory}

\subsection{Correspondence between the flow and metric perturbations}
\label{app:metric}

We revisit the linear response theory for the charge density induced in the magnetovortical matter, and clarify the correspondence between our result in the main text and that in Ref.~\cite{Hattori:2016njk}. 
Within the linear-response regime, the current induced by the Hamiltonian density $ H_\mathrm{ext}$ is given as \cite{Kapusta:2006pm}
\begin{equation}
    \delta j^\mu (t,\bx)
    = - \rmi \int_{-\infty}^t \rmd t' \int \rmd^3x'
    \langle [ j^\mu(t,\bx), H_\mathrm{ext} (t',\bx')] \rangle
    , 
\label{eq:linear-response-current}
\end{equation}
where the angle brackets represent the thermal average.
The perturbation Hamiltonian density is identified as follows.

We first check the equivalence between a flow perturbation and a metric perturbation used in Ref.~\cite{Hattori:2016njk}. 
The Hamiltonian for the rest fluid $\mathring{u}^\mu = (1,\boldsymbol{0})$ is given as 
\begin{equation}
 \mathring{H} 
 = T^{0\mu} \mathring{u}_\mu = T^{00}.
\end{equation}
Let us consider a nontrivial background fluid velocity $u^\mu = (1, \boldsymbol{v})$, i.e, $u_\mu = (1, -\boldsymbol{v})$, where we ignore the terms of $O(|\bv|^2)$.
Then, the Hamiltonian at the comoving frame reads 
\begin{equation}
 H = T^{0\mu} u_\mu = 
 \mathring{H} - T^{0i} v^i .
 \label{eq:H-flow}
\end{equation}
In particular, let us consider a flow $\boldsymbol{v}=\boldsymbol{\Omega}\times\boldsymbol{x}$ with a constant $\boldsymbol{\Omega}$ that realizes the rotating system with the angular velocity $\frac{1}{2}\boldsymbol{\nabla}\times\boldsymbol{v} = \boldsymbol{\Omega}$. 
Then, the above Hamiltonian becomes
\begin{equation}
 H = 
 \mathring{H} - \boldsymbol{\Omega}\cdot(\boldsymbol{x}\times\boldsymbol{P}), 
    \label{eq:H-Omega}
\end{equation}
where $ P^i=(\boldsymbol{P})^i = T^{0i}$.

This Hamiltonian is mimicked by a background geometry.
That is, we consider a static curved spacetime quantified by the metric tensor $g_{\mu\nu} = \eta_{\mu\nu} + h_{\mu\nu}$, where the Minkowskian metric is given as $\eta_{\mu\nu} = \mathrm{diag}(1,-1,-1,-1)$. 
We suppose that the metric perturbation $h_{\mu\nu}$ has nonzero components only in $h_{0i}$.
The Hamiltonian for the rest fluid in this geometry is specified by $u^\mu = (1,\boldsymbol{0})$, i.e., $u_\mu = \mathring{u}_\mu +  h_{0\mu}$ and reads
\begin{equation}
 H = T^{0\mu} u_\mu = 
 \mathring{H} + T^{0i} h_{0i}.
 \label{eq:H-metric}
\end{equation}
According to Eqs.~\eqref{eq:H-flow} and \eqref{eq:H-metric}, 
the perturbation Hamiltonian is now identified as
\begin{equation}
H_\mathrm{ext}=-T^{0i}v^{i}=T^{0i}h_{0i}.
\end{equation}
This implies the correspondence between the perturbations 
\begin{equation}
\label{eq:v-g}
    v^i = -h_{0i} ,
\end{equation} 
and thus that 
\begin{equation}
\label{eq:Omega}
 \Omega^i = -\frac{1}{2}\epsilon^{ijk} \partial_j h_{0k} ,
\end{equation}
where the Levi-Civita symbol is normalized by $\epsilon^{123}=-\epsilon_{123}=1$.
The same expression of the vorticity is shown in Ref.~\cite{Landau:classical} and required to reproduce the chiral vortical effect in Ref.~\cite{Hayata:2020sqz} with the correct sign. 
However, above Eq.~(8) of Ref.~\cite{Hattori:2016njk}, the relation between the perturbations has an opposite sign.

\subsection{Linear response under a metric perturbation}

Let us analyze the induced charge under the metric perturbation. 
In the following, all indices can be raised and lowered by $\eta_{\mu\nu}$, granted that we consider the linear response to the metric perturbation. 
The linear response theory \eqref{eq:linear-response-current} tells that, under a weak metric perturbation \eqref{eq:H-metric},  the induced current is expressed with the retarded correlator as 
\begin{equation}
\begin{split}
\label{eq:GR_JT}
 \delta j^\mu(q) = \frac{1}{2} G^{\mu\nu\rho}_\R(q) h_{\nu\rho}(q) ,
 \quad
 G^{\mu\nu\rho}_\R (x-x') = -\rmi \theta(t-t') \langle [j^\mu(x),T^{\nu\rho}(x')] \rangle ,
\end{split}
\end{equation}
where we defined $q^\mu = (\omega,\bq)$ and the Fourier transform 
\begin{equation}
 f (x) = \int \frac{\rmd^4 q}{(2\pi)^4}\, \rme^{-\rmi q\cdot x} f(q).
\end{equation}
The factor of $1/2$ is attached in front of the correlator, in order to cancel a doubled contribution from the symmetrized contraction. 
The microscopic definitions of the current and the energy-momentum tensor are given by 
\begin{equation}
j^\mu (x) = \bar{\psi} (x) \gamma^\mu \psi (x),
\quad  T^{0i} (x) = \frac{\rmi}{2}\bar{\psi} (x) (\gamma^0 D^i + \gamma^i D^0)\psi(x) .
\end{equation}
The covariant derivative $ D_\mu = \partial_\mu + \rmi eA_\mu$ contains $A_\mu$ for a constant magnetic field. 
Based on the analytic continuation, we relate the retarded correlator to the imaginary-time correlator as 
\begin{equation}
\label{eq:GE_AC}
  G^{\mu\nu\rho}_\E (q_0\to \omega +\rmi\delta,\bq) = G_\R^{\mu\nu\rho}(q) ,
 \quad
 G_\E^{\mu\nu\rho}(x-x') = -\langle T_\tau j^\mu (x) T^{\nu\rho}(x')\rangle ,
\end{equation}
where $\delta$ is a positive infinitesimal number for the analytic continuation and $T_\tau$ is the imaginary-time ordering. 
The momentum in the imaginary-time correlator is $q^\mu = (q_0,\bq)$ with the bosonic Matsubara frequency $q_0=2\rmi\pi T m$ ($m$ is an integer). 
Focusing on the charge density, we have 
\begin{equation}
\label{eq:GE_J0T}
  \delta j^0 = 
  h_{0k}(q) \lim_{\omega\to 0}G_\E^{00k}(q_0\to \omega +\rmi\delta,\bq) .
\end{equation}
The limit of $\omega\to 0$ is required in order to obtain a static transport coefficient.

We compute the above correlator in the lowest Landau level (LLL) limit at the one-loop order. 
One should start with the coordinate-space representation 
\begin{equation}
\begin{split}
\label{eq:GE00k}
G_\E^{00k}(x-x')
&= (-1)\cdot (-1)\frac{\rmi}{2}\tr\Bigl[\calP_+\gamma^0 S_\LLL(x,x') (\gamma^0 D^k_{x'}+\gamma^k D^0_{x'}) S_\LLL(x',x) \Bigr] \\
&= \frac{\rmi}{2}\tr\Bigl[\calP_+\gamma^0 S_\LLL(x,x') \gamma^0 D^k_{x'}S_\LLL(x',x) \Bigr] ,
\end{split}
\end{equation}
where the two minus signs are due to the definition of $G_\E^{\mu\nu\rho}$ in Eq.~\eqref{eq:GE_AC} and the fermion loop.
The LLL propagator $S_\LLL(x',x)$ is given by
\begin{equation}
 S_\LLL(x',x)
 = \langle\bar\psi_\LLL(x') \psi_\LLL(x) \rangle
 = \rme^{\rmi \phi (x',x)}\tilde{S}_\LLL(x'-x)
\end{equation}
with $\psi_\LLL(x)$ is the field operator for the Dirac fermion in the LLL. 
The translational invariance is broken by the Schwinger phase 
\begin{equation}
  \phi (x',x) = -e\int_x^{x'} \rmd z^\mu \biggl[ A_\mu (z) + \frac{1}{2}F_{\mu\nu}(z^\nu-x^\nu) \biggr],
\end{equation}
which fulfills
\begin{equation}
\begin{split}
 D^\mu_{x'} S_\LLL(x',x)
 &= \rme^{\rmi \phi (x',x)}\biggl[
 	\partial^\mu_{x'}
 	-\frac{\rmi e}{2}F^{\mu\nu}({x}'_\nu-x_\nu)
   \biggr]\tilde{S}_\LLL(x'-x) .
\end{split}
\end{equation}
The translation-invariant part of the imaginary-time propagator is given as 
\begin{equation}
\label{eq:SLLL}
 \tilde S_\LLL(p) = 2\,\rme^{-|p_\perp|^2/|eB|} 
 S_{1+1} (p_\parallel),
 \quad
 S_{1+1}(p_\parallel) = \frac{\calP_+}{\gamma^\mu_\parallel p_{\mu\parallel}}
 = \frac{\calP_+}{\gamma^0 p_0 + \gamma^3 p_3 } ,
\end{equation}
where $p^\mu = (p_0,\bp)$, $p_0=\rmi\pi T(2n-1)$, $\bp = (p^1,p^2,p^3)$,  $p^\mu_\parallel = (p^0,0,0,p^3)$, $p^\mu_\perp = (0,p^1,p^2,0)$ and $\calP_+ = (1+\rmi \gamma^1\gamma^2)/2$.
The field strength is now specified as $ F^{21} = \partial^2 A^1 - \partial^1 A^2 = \partial_1 A^2 -\partial_2 A^1 = B$, namely,
\begin{equation}
 F^{kl} = -\epsilon^{kl}_\perp B.
\end{equation}
The transverse Levi-Civita symbol is defined by $\epsilon_\perp^{12} = - \epsilon_\perp^{21} = 1$, i.e., $\epsilon^{jk}_\perp = \epsilon^{3jk}$.

From the above preparation, the correlation function is written as
\begin{equation}
\begin{split}
\label{eq:GE00k_FT}
G_\E^{00k}(x-x')
&= \frac{\rmi}{2}\tr
 \Bigl[\calP_+
 	\gamma^0 \tilde{S}_\LLL(x-x')\gamma^0 
 	\Bigl(\partial^k_{x'}
 	+\rmi (eB/2)\epsilon_\perp^{kl}({x}'_l-x_l)\Bigr)\tilde{S}_\LLL(x'-x)
 \Bigr] \\
 &=\frac{\rmi}{2}\,
	T\sum_{n} \int \frac{\rmd^3 p}{(2\pi)^3}\, T\sum_{l} \int \frac{\rmd^3 q}{(2\pi)^3}\, \rme^{-\rmi (q-p)\cdot(x-x')}
 \tr
 \Bigl[\calP_+
 	\gamma^0 \tilde S_\LLL(q)\gamma^0 (-\rmi p^k + (eB/2)\epsilon_\perp^{kl} \partial^p_l) \tilde S_\LLL(p)
 \Bigr] \\
 &= \frac{1}{2}\,
	T\sum_{n} \int \frac{\rmd^3 p}{(2\pi)^3}\, T\sum_{l} \int \frac{\rmd^3 q}{(2\pi)^3}\, \rme^{-\rmi q\cdot(x-x')}
 \tr
 \Bigl[\calP_+
 	\gamma^0 \tilde S_\LLL(p+q)\gamma^0 (p^k +\rmi (eB/2)\epsilon_\perp^{kl} \partial^p_l) \tilde S_\LLL(p)
 \Bigr] .
\end{split}
\end{equation}
The last line is obtained with the change of the variable $q-p\to q$. 
Here, we define $\partial_p^j  = \partial/\partial p_j 
= - \partial/\partial p^j $, and accordingly $ \partial^p_j =- \partial_p^j $. 
From Eq.~\eqref{eq:GE00k_FT}, the correlation function in the momentum space reads
\begin{equation}
\begin{split}
 G_\E^{00k}(q)
 &= \frac{1}{2}\,
	T\sum_{m} \int \frac{\rmd^3 p}{(2\pi)^3}
 \tr
 \Bigl[\calP_+
 	\gamma^0 \tilde S_\LLL(p+q)\gamma^0 (p^k +\rmi (eB/2)\epsilon_\perp^{kl} \partial^p_l) \tilde S_\LLL(p)
 \Bigr].
\end{split}
\end{equation}
Hereafter, we focus only on $k=1,2$, that is, the transverse component of $G^{00k}_{\E\perp}(q)$.
The sign in front of $ eB$ is opposite to that in Eq.~(15) of Ref.~\cite{Hattori:2016njk} for $ k=2$.

The phase space integral in $G^{00k}_{\E\perp}(q)$ is decomposed into those over the transverse and longitudinal momenta, as follows:
\begin{equation}
 G^{00k}_{\E\perp}(q) = I^k_\perp(q_\perp) I_\parallel(q) ,
\end{equation}
\begin{equation}
I^k_\perp(q_\perp) = 2 \int\frac{\rmd^2 p_\perp}{(2\pi)^2} 
(p_\perp^k +\rmi s \epsilon_\perp^{lk} p_\perp^l)
\, \rme^{-(|p_\perp+q_\perp|^2+|p_\perp|^2)/|eB|} ,
\end{equation}
\begin{equation}
\begin{split}
 I_\parallel(q_\parallel)
 = T\sum_n\int \frac{\rmd p_z}{2\pi} \, \tr
 \Bigl[\calP_+
 	\gamma^0 S_{1+1}(p_\parallel+q_\parallel)\gamma^0 S_{1+1}(p_\parallel)
 \Bigr] , \label{eq:I_para-def}
\end{split}
\end{equation}
where $s = {\rm sgn}(eB) $.
The former integration can be performed as 
\begin{equation}
\begin{split}
\label{eq:Ik_perp}
I^k_\perp(q_\perp)
&= 2  \rme^{-|\bq_\perp|^2/(2|eB|)} 
\int \frac{\rmd^2p_\perp}{(2\pi)^2}  
(p_\perp^k +\rmi s\epsilon_\perp^{lk} p_\perp^l)
\rme^{- 2 |\bp_\perp + \bq_\perp/2|^2/|eB|} \\
&= -\rme^{-|\bq_\perp|^2/(2|eB|)} 
(q_\perp^k +\rmi s\epsilon_\perp^{lk} q_\perp^l) 
\int \frac{\rmd^2p'_\perp}{(2\pi)^2} 
\rme^{- 2 |\bp_\perp'|^2/|eB| } \\
&= - \frac{|eB|}{8\pi}  \rme^{-|\bq_\perp|^2/(2|eB|)} 
(q_\perp^k +\rmi s\epsilon_\perp^{lk} q_\perp^l) .
\end{split}
\end{equation}
For the latter integral, one can replace the Matsubara summation 
by a contour integral.
By the use of identities $\tanh z = \pm [ 1 - 2 /( \rme^{\pm 2z} + 1)] $, 
one can separate the vacuum and thermal parts as $I_\parallel = I_\parallel^\mathrm{vac} + I_\parallel^\mathrm{mat}$ with
\begin{equation}
\begin{split}
\label{eq:vacuum-thermal}
 I_\parallel^\mathrm{vac}(q_\parallel) 
 &= \int\frac{\rmd p_z}{2\pi} 
 \int_{-\rmi \infty }^{\rmi \infty} \frac{\rmd p_0 }{2\pi\rmi}\,
 \tr\biggl[ \calP_+\gamma^0 \frac{1}{\gamma^\mu_\parallel(p_{\mu\parallel}+q_{\mu\parallel})} 	\gamma^0 \frac{1}{\gamma^\nu_\parallel p_{\nu\parallel}} \biggr] ,\\
 I_\parallel^\mathrm{mat}(q_\parallel) 
 &= -\int\frac{\rmd p_z}{2\pi} 
 \sum_{\eta=\pm}\int_{C_\eta} \frac{\rmd p_0 }{2\pi\rmi} n_\mathrm{F}(\eta p_0) \,
 \tr\biggl[ \calP_+\gamma^0 \frac{1}{\gamma^\mu_\parallel(p_{\mu\parallel}+q_{\mu\parallel})} 	\gamma^0 \frac{1}{\gamma^\nu_\parallel p_{\nu\parallel}} \biggr] ,
\end{split}
\end{equation}
where we defined $p_z = p^3$ and the thermal distribution function $n_\mathrm{F}(p_0) =1/[ \rme^{\beta p_0 } + 1]$.
The contours $C_\pm$ go along $-\rmi\infty \pm\varepsilon \to \rmi\infty \pm\varepsilon$ with a positive infinitesimal number $\varepsilon$, and pick up the residues at the positive- and negative-energy poles, respectively. 
 
It is here important to notice that the thermal contribution $I_\parallel^\mathrm{mat}(q_\parallel)$ vanishes totally, as shown in the following.
First, we compute the spinor trace as $\tr[\calP_+\gamma^0 \gamma^\mu_\parallel \gamma^0 \gamma^\nu_\parallel]= 2(-\eta^{\mu\nu}_\parallel + 2\eta^{\mu 0}_\parallel \eta^{\nu 0}_\parallel)$ with $\eta^{\mu\nu}_\parallel$ being the longitudinal components of the Minkowskian metric tensor.
The contour integral is then evaluated with the residues as 
\begin{equation}
\begin{split}
\label{eq:thermal}
I_\parallel^\mathrm{mat}
&=\sum_{\eta=\pm}\eta\!\int\!\frac{\rmd p_z}{2\pi} \biggl[
	 n_\mathrm{F}(|p_z|)\frac{\eta p_z(\eta p_z+q_0)+p_z(p_z+q_z)}{2\eta p_z[(\eta p_z+q_0)^2-(p_z+q_z)^2]}
 +n_\mathrm{F}(|p_z+q_z|)\frac{(\eta(p_z+q_z)-q_0)\eta(p_z+q_z)+p_z(p_z+q_z)}{2[(\eta(p_z+q_z)-q_0)^2-p_z^2]\eta(p_z+q_z)}  
 \biggr]\\
&= \frac{1}{2}\sum_{\eta=\pm}\frac{1}{q_z-\eta q_0} \cdot
\int\frac{\rmd p_z}{2\pi} \Bigl[ n_\mathrm{F}(|p_z+q_z|)-n_\mathrm{F}(|p_z|) \Bigr],
\end{split}
\end{equation}
where we used $\rme^{\eta\beta q_0} = \rme^{2\rmi \pi m} = 1$ for $q_0 = 2\rmi\pi T m$.
Since the above $p_z$-integrals are convergent, we can freely shift the integral variable. 
This manipulation leads to the cancellation between the two integrals in Eq.~\eqref{eq:thermal}, and thus the conclusion that $I_\parallel^\mathrm{mat}(q_\parallel)=0$.

The remaining piece is the vacuum contribution. 
This can be computed with the standard Feynman-parameter trick as~\cite{Dolan:1973qd,*Baier:1991gg,*Fukushima:2015wck,Hattori:2022uzp} 
\begin{equation}
\begin{split}
I_\parallel(q_\parallel) 
= I^\mathrm{vac}_\parallel(q_\parallel) 
&= -\frac{1}{\rmi}\int\frac{\rmd^2 p_\parallel}{(2\pi)^2} 
 \tr\biggl[ \calP_+\gamma^0 \frac{\rmi}{\gamma^\mu_\parallel(p_{\mu\parallel}+q_{\mu\parallel})} 	\gamma^0 \frac{\rmi}{\gamma^\nu_\parallel p_{\nu\parallel}} \biggr] \\
&=- \frac{1}{\pi}\cdot \frac{ q^2_\parallel - q_0^2 }{q_\parallel^2} 
= \frac{1}{\pi}\cdot \frac{ q_z^2 }{ q_0^2 - q_z^2} ,
\label{eq:I_parallel-0}
\end{split}
\end{equation} 
where we rotate the contour of the $p_0$-integral from $-\rmi \infty \to \rmi \infty$ to $-\infty \to \infty$. 
This result is equal to the polarization function in the $(1+1)$-dimensional QED {\it divided} by a factor of $(-\rmi e)^2$ \cite{Hattori:2022uzp}. 
However, Eq.~(19) of Ref.~\cite{Hattori:2016njk} has an opposite sign.

Performing the analytic continuation and taking the static limit, we have 
\begin{equation}
\label{eq:I_parallel}
\lim_{\omega\to0} I_\parallel(q_0\to \omega + \rmi\delta,q_z) = - \frac{1}{\pi} .
\end{equation}
Finally, the correlation function reads
\begin{equation}
 \lim_{\omega\to 0}G_\E^{00k}(q_0\to \omega +\rmi\delta,\bq)
 = \biggl(-\frac{1}{\pi}\biggr)\cdot \biggl(-\frac{eB}{8\pi}\biggr) (q_\perp^k + \rmi \epsilon^{3jk} q_\perp^j)
 + O(|q_\perp|/\sqrt{|eB|}).
\end{equation}
Therefore, the vorticity-induced term reads
\begin{equation}
\label{eq:J_vortical}
 \delta j_\mathrm{vortical}^0 = \frac{eB}{8\pi^2} \epsilon^{3jk}  \partial_j h_{0k}
 = -\frac{eB \Omega}{4\pi^2}
\end{equation}
with $\Omega = -\epsilon^{3jk} \partial_j h_{0k}/2$.
This is consistent with the isothermal transport coefficient that is derived from the partition function in the main text.
In summary, the three opposite signs in Eqs.~\eqref{eq:Omega}, \eqref{eq:GE00k_FT}, and~\eqref{eq:I_parallel-0} explain why Eq.~\eqref{eq:J_vortical} has the extra minus sign compared with the main result in Ref.~\cite{Hattori:2016njk}. 
This concludes that the linear-response analysis in the strong magnetic field contains both the orbital and spin contributions.

\subsection{Linear response under a perturbative rotation}
We show that the same charge density is reproduced from the linear-response theory in terms of a weak angular velocity $\Omega$ based on the correspondence in Eq.~\eqref{eq:v-g}. 
The kinetic angular momentum operator explicitly appears in this analysis. 

Let us consider the charge density $\delta j^0 = \lambda\Omega$ generated by the perturbation Hamiltonian~\eqref{eq:H-Omega}, i.e., 
\begin{equation}
 H_\mathrm{ext}(x) = -\Omega J_\kin (x),
 \quad
 J_\kin = \frac{\rmi}{2}\epsilon^{3jk}\,x^j \bar\psi (\gamma^0 D^k + \gamma^k D^0)\psi.
\end{equation}
As in the above analysis, the transport coefficient $\lambda$ is computed from the imaginary-time correlator as 
\begin{equation}
\label{eq:lambda}
 \lambda = \lim_{\bq \to \boldsymbol{0}}\Bigl[\lim_{\omega \to 0}G_\E(q_0 \to \omega+\rmi\delta , \bq)\Bigr],
 \quad
 G_\E (x,y)= -\Bigl\langle T_\tau j^0 (x) \Bigl(-J_\kin (y)\Bigr)\Bigr\rangle.
\end{equation}

We compute the above correlator in the LLL limit at the one-loop order. 
A similar correlator has been directly evaluated with the covariant derivative $D_\mu$ in the above and in Ref.~\cite{Hattori:2016njk}. 
Here, it is more useful to evaluate the covariant derivative as the kinetic angular momentum operator. 
That is, we can utilize the relation 
\begin{equation}
 \frac{\rmi}{2}\epsilon^{3jk}\,x^j \bar\psi_\LLL \gamma^0 D^k\psi_\LLL
 = \frac{1}{2} \bar\psi_\LLL \gamma^0 L_\kin \psi_\LLL 
 = -\frac{s}{2} \bar\psi_\LLL \gamma^0 \psi_\LLL.
 \label{eq:L_kin-LLL}
\end{equation}
The kinetic angular momentum is evaluated as $L_\kin \psi_\LLL = -s\psi_\LLL + \text{(higher LL)}$ based on Eq.~(2) in the main text. 
Thus, the angular momentum is expressed with the charge density:
\begin{equation}
\begin{split}
J_\kin (y)
&= -\frac{s}{2}  j^0(y)
+\frac{\rmi}{2}\epsilon^{3jk}\,x^j \bar\psi \gamma^k D^0\psi.
\end{split}
\end{equation}
Since the second term should vanish due to the spinor trace, the correlator $G_\E$ takes the same form as the density-density correlator:
\begin{equation}
\begin{split}
 G_\E(x,y)
 &= (-1)\cdot\biggl(-\frac{s}{2}\biggr)\tr\Bigl[\calP_+\gamma^0 S_\LLL(x,y) \gamma^0 S_\LLL (y,x)\Bigr],
\end{split}
\end{equation}
where the additional minus sign is from the fermion loop.
The momentum representation of the correlator reads
\begin{equation}
\begin{split}
 G_\E(q)
 &= \frac{s}{2}\,T
	\sum_{n}\int\frac{\rmd^3 p}{(2\pi)^3}
 \tr
 \Bigl[\calP_+
 	\gamma^0 \tilde S_\LLL(p+q)\gamma^0 \tilde S_\LLL(p)
 \Bigr] .
\end{split}
\end{equation}
One can further decompose the integral as $G_\E(q)=\tilde{I}_\perp(q_\perp) I_\parallel(q_\parallel)$ with the same $I_\parallel(q_\parallel)$ as in Eq.~\eqref{eq:I_para-def}. 
The transverse part $\tilde{I}_\perp(q_\perp) $ is slightly different from Eq.~\eqref{eq:Ik_perp}, but is readily evaluated in the same manner as 
\begin{equation}
\begin{split}
\label{eq:I_perp}
 \tilde{I}_\perp (q_\perp)
 = 2s\int \frac{\rmd^2 p_\perp}{(2\pi)^2}\,\rme^{-(|p_\perp|^2+|p_\perp+q_\perp|^2)/|eB|}
 = \frac{eB}{4\pi} \rme^{-|q_\perp|^2/|eB|}.
\end{split}
\end{equation}
Plugging Eqs.~\eqref{eq:I_parallel} and~\eqref{eq:I_perp} into Eq.~\eqref{eq:lambda}, we finally arrive at
\begin{equation}
 \lambda = -\frac{eB}{4\pi^2}.
\end{equation}
This result again agrees with the induced charge density derived in the main text. 
The origin of the negative sign is clear from Eq.~\eqref{eq:L_kin-LLL}. 
We reiterate that, according to the correspondence in Eq.~\eqref{eq:v-g}, this analysis is equivalent to the analysis leading to Eq.~\eqref{eq:J_vortical} with the metric perturbation. 
%

\end{document}